\documentclass{article}

\usepackage{nicematrix}

\usepackage{framed}

\usepackage{lipsum}

\usepackage{tikz, pgf, pgfplots}
\usetikzlibrary
	{
		arrows, automata, chains, datavisualization.formats.functions, fadings, fit, positioning, quantikz, quotes, shapes
	}
\usepackage{tikzpeople}
\usepackage{fontawesome5}
\newcounter{mathseed}
\pgfmathsetseed{\arabic{mathseed}}
\pgfdeclarelayer{background}
\pgfsetlayers{background,main}
\pgfdeclaredecoration{irregular fractal line}{init}
{
	\state{init}[width=\pgfdecoratedinputsegmentremainingdistance]
	{
		\pgfpathlineto{\pgfpoint{random*\pgfdecoratedinputsegmentremainingdistance}{(random*\pgfdecorationsegmentamplitude-0.02)*\pgfdecoratedinputsegmentremainingdistance}}
		\pgfpathlineto{\pgfpoint{\pgfdecoratedinputsegmentremainingdistance}{0pt}}
	}
}
\tikzset
{
	paper/.style =
	{
		draw = MyDarkBlue!10, blur shadow, every shadow/.style = { opacity = 1, MyDarkBlue }, shading = bilinear interpolation, lower left = MyDarkBlue!10, upper left = MyDarkBlue!5, upper right = GreenTeal!75, lower right = MyDarkBlue!5, fill=none
	},
	irregular cloudy border/.style =
	{
		decoration = { irregular fractal line, amplitude = 0.2 }, decorate,
	},
	irregular spiky border/.style =
	{
		decoration = { irregular fractal line, amplitude = -0.2 }, decorate,
	},
	ragged border/.style =
	{
		decoration = {random steps, segment length = 7mm, amplitude = 2mm }, decorate
	}
}
\def\tornpaper#1{%
	\ifthenelse{\isodd{\value{mathseed}}}
	{%
		\tikz
		{
			\node[inner sep = 1em] (A) {#1};		
			\begin{pgfonlayer}{background}			
				\fill[paper]						
				\pgfextra{\pgfmathsetseed{\arabic{mathseed}}\addtocounter{mathseed}{1}}%
				{decorate[irregular cloudy border]{decorate{decorate{decorate{decorate[ragged border]{
										(A.north west) -- (A.north east)
				}}}}}}
				-- (A.south east)
				\pgfextra{\pgfmathsetseed{\arabic{mathseed}}}%
				{decorate[irregular spiky border]{decorate{decorate{decorate{decorate[ragged border]{
										-- (A.south west)
				}}}}}}
				-- (A.north west);
			\end{pgfonlayer}
		}
	}
	{%
		\tikz{
			\node[inner sep=1em] (A) {#1};  
			\begin{pgfonlayer}{background}  
				\fill[paper] 
				\pgfextra{\pgfmathsetseed{\arabic{mathseed}}\addtocounter{mathseed}{1}}%
				{decorate[irregular spiky border]{decorate{decorate{decorate{decorate[ragged border]{
										(A.north east) -- (A.north west)
				}}}}}}
				-- (A.south west)
				\pgfextra{\pgfmathsetseed{\arabic{mathseed}}}%
				{decorate[irregular cloudy border]{decorate{decorate{decorate{decorate[ragged border]{
										-- (A.south east)
				}}}}}}
				-- (A.north east);
		\end{pgfonlayer}}
	}
}

\usepackage{amsthm}
\usepackage{amssymb}
\usepackage[bb = boondox]{mathalfa}
\usepackage{dsfont}
\usepackage{mathtools}
\usepackage{multicol}
\usepackage{braket}
\usepackage{physics}
\numberwithin{equation}{section}

\usepackage{yquant}

\usepackage[a4paper, left = 2.5cm, right = 2.5 cm, top = 2.5cm, bottom = 3.0 cm]{geometry}

\definecolor{MyLightRed}{RGB}{244, 213, 245}
\definecolor{WordRed}{RGB}{255, 0, 102}
\definecolor{RedDarkLightest}{HTML}{ff0088}
\definecolor{RedDarkLight}{HTML}{ea005f}
\definecolor{RedPurple}{HTML}{aa007f}
\definecolor{Purple}{HTML}{911146}
\definecolor{WordLightGreen}{RGB}{140, 214, 192}
\definecolor{WordGreen}{RGB}{0, 176, 80}
\definecolor{GreenLightest}{HTML}{00ffa0}
\definecolor{GreenLighter1}{HTML}{00b383}
\definecolor{GreenLighter2}{HTML}{00aa7f}
\definecolor{GreenDark}{HTML}{225522}
\definecolor{GreenTeal}{HTML}{008080}
\definecolor{WordIceBlue}{RGB}{223, 227, 229}
\definecolor{MyVeryLightBlue}{RGB}{211, 245, 247}
\definecolor{WordBlueVeryLight}{RGB}{0, 176, 240}
\definecolor{WordBlueLight}{RGB}{0, 112, 192}
\definecolor{WordBlueDark}{RGB}{46, 116, 181}
\definecolor{WordBlueDarker}{RGB}{31, 78, 121}
\definecolor{WordBlueDarker25}{RGB}{54, 96, 146}
\definecolor{WordBlueDarker50}{RGB}{36, 64, 98}
\definecolor{WordBlueDarkest}{RGB}{0, 32, 96}
\definecolor{WordBlue}{RGB}{19, 65, 99}
\definecolor{MyBlue}{RGB}{0, 64, 128}
\definecolor{MyDarkBlue}{RGB}{0, 51, 102}
\definecolor{BlueVeryDark}{HTML}{222255}
\definecolor{MagentaVeryLight}{RGB}{178, 162, 201}
\definecolor{MagentaLighter}{RGB}{161, 106, 221}
\definecolor{MagentaLight}{RGB}{128, 100, 162}
\definecolor{MagentaDark}{RGB}{106, 65, 152}
\definecolor{MagentaVeryDark}{RGB}{97, 75, 128}
\definecolor{WordAquaLighter80}{RGB}{218, 238, 243}
\definecolor{WordAquaLighter60}{RGB}{183, 222, 232}
\definecolor{WordAquaLighter40}{RGB}{146, 205, 220}
\definecolor{WordAquaDarker25}{RGB}{49, 134, 155}
\definecolor{WordAquaDarker50}{RGB}{33, 89, 103}
\definecolor{WordVeryLightTeal}{RGB}{223, 236, 235}
\definecolor{WordLightTeal}{RGB}{160, 199, 197}
\definecolor{WordDarkTealLighter80}{RGB}{207, 223, 234}
\definecolor{WordDarkTeal}{RGB}{72, 123, 119}
\definecolor{WordDarkerTeal}{RGB}{48, 82, 80}
\definecolor{WordTurquoiseLighter80}{RGB}{209, 238, 249}
\definecolor{Brown}{HTML}{666633}

\usepackage{xcolor}
\usepackage{varwidth}
\usepackage[breakable, skins, theorems]{tcolorbox}
\usepackage{graphicx}
\usepackage{hyperref}
\hypersetup
{
	colorlinks,
	linkcolor = {WordRed!75!black},
	citecolor = {WordBlueDarker25},
	urlcolor = {WordAquaDarker50}
}
\usepackage{colortbl}
\usepackage{float}
\usepackage{cellspace}
\usepackage{makecell}
\usepackage[boxed, ruled, vlined]{algorithm2e}
\usepackage{appendix}
\usepackage{multirow}
\newcommand{\orcidicon}[1]{\href{https://orcid.org/#1}{\includegraphics[height=\fontcharht\font`\B]{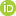}}}

\newtheorem{example}{Example}[section]

\title
	{
		An entanglement-based protocol for simultaneous reciprocal information exchange between 2 players
	}

\author
	{
		Theodore Andronikos$^1$\orcidicon{0000-0002-3741-1271}
		and
		Alla Sirokofskich$^2$\\
		\\
		$^1$ \ Department of Informatics, Ionian University, \\
		7 Tsirigoti Square, 49100 Corfu, Greece; \\
		andronikos@ionio.gr \\
		$^2$ \ Department of History and Philosophy of Sciences, \\
		National and Kapodistrian University of Athens, \\
		Athens 15771, Greece; \\
		asirokof@math.uoa.gr
	}

\begin{document}

\maketitle

\begin{abstract}
	Let us consider a situation where two information brokers, whose currency is, of course, information, need to reciprocally exchange information. The two brokers, being somewhat distrustful, would like a third, mutually trusted, entity to be involved in the exchange process so as to guarantee the successful completion of the transaction, and also verify that it indeed took place. Can this be done in such a way that both brokers receive their information simultaneously and securely, and without the trusted intermediary ending up knowing the exchanged information? This work presents and rigorously analyzes a new quantum entanglement-based protocol that provides a solution to the above problem. The proposed protocol is aptly named entanglement-based reciprocal simultaneous information exchange protocol. Its security is ultimately based on the assumption of the existence of a third trusted party. Although, the reciprocal information flow is between our two information brokers, the third entity plays a crucial role in mediating this process, being a guarantor and a verifier. The phenomenon of quantum entanglement is the cornerstone of this protocol, as it makes possible its implementation even when all entities are spatially separated, and ensuring that, upon completion, the trusted third party remains oblivious of the actual information that was exchanged.
\end{abstract}
\section{Introduction} \label{sec:Introduction}

Living in an era where privacy and security are fundamental and inherent rights in everyone's professional and social life, has undoubtedly motivated the enormous research effort to design and implement robust security algorithms, techniques, and protocols. It would seem that there are at least two major research directions aspiring to achieve this goal.

The first, known by the umbrella term post-quantum cryptography \cite{chen2016report, alagic2019status, alagic2020status, alagic2022status}, could be perceived as the natural evolution of our present situation, where security relies on carefully chosen computationally hard problems. The efficiency and reliability of this approach has been practically confirmed thus far, and offers the additional advantage that implementation-wise it is compatible with the existing infrastructure.

There is however reason for concern, as the computational difficult has not actually been mathematically proven, but is rather empirically accepted because of the absence of efficient algorithms. Probably, even more worrying is the fact that when one steps out of the confines of conventional, or classical, computation, discovers that quantum algorithms have been developed by Peter Shor and Lov Grover \cite{Shor1994, Grover1996}, that may compromise the security afforded by conventional means. The second direction, quantum cryptography, advocates the embrace of the unconventional, in particular the quantum, realm and relies on the laws of nature, at least as we understand them today, to achieve uncompromising security.

On this front, there are many reasons to be optimistic lately, with signs of accelerated progress. The impressive breakthrough of the $100$ qubit barrier by IBM's $127$ qubit processor Eagle \cite{IBMEagle} has been, almost immediately, followed by the more recent $433$ qubit quantum processor named Osprey \cite{IBMOsprey}. Therefore, it is now more important than ever before to address the problem of secure communication \cite{chamola2021information}, and quantum cryptography seems the most promising approach for the job. Clever use of the fundamental properties of quantum mechanics offers undisputed advantages that not only they guarantee the adequate protection of critical information, but also ensure the efficient and secure transmission of information, e.g., via the use of entanglement, as first suggested by Ekert \cite{Ekert1991}. In his 1991 influential paper, Ekert proved that key distribution is possible with the use of EPR pairs. Immediately afterwards, research in this field produced a plethora of entanglement-based protocols for quantum key distribution \cite{Bennett1992, Gisin2004, inoue2002differential, guan2015experimental, waks2006security, Ampatzis2021, Ampatzis2022}. Quantum cryptographic protocols have the potential to enhance the security not only of established applications, but also of new and emerging technologies, such as cloud computing, cloud storage \cite{attasena2017secret, ermakova2013secret} or blockchain \cite{cha2021blockchain}. Considerable progress has been recorded in the development of algorithm and protocols \cite{karlsson1999quantum, smith2000quantum, gottesman2000theory, fortescue2012reducing, qin2020hierarchical, senthoor2022theory}, accompanied by experimental demonstrations in real-life situations \cite{fu2015long, wu2020passive, grice2019quantum, gu2021secure}.

This work studies the following problem within the broader context of quantum information and cryptography. We envision a setting where information is the real currency, and in such a setting, two information brokers want to reciprocally exchange information. In their profession, trust can not be taken for granted. Hence, they would like a third party that is mutually trusted by both, to mediate the whole process and guaranteed the honest implementation of the agreed upon protocol. This third party must have a pivotal role, in the sense that without its participation it will be impossible to complete the exchange. Another important benefit of the presence of the trusted intermediary is that it can serve as a referee to verify that this transaction took place, if such a need arises in the future, e.g., if one or both of the information brokers ever need such a proof.
This scheme should work even when all parties involved are spatially separated, and, most importantly, designed so carefully that the trusted intermediary does not end up in possession of the exchanged information. In the rest of this paper we present a quantum protocol that provides a solution to this problem, satisfying all the previously set requirements, in the form of a game.

It is almost some kind of tradition to use games in quantum cryptography; from the very beginning  \cite{Bennett1984} to the more recent \cite{Ampatzis2021, Ampatzis2022}. The pedagogical aspect of games often makes expositions of difficult and technical concepts much more accessible, even entertaining. Quantum games, ever since their initial inception in 1999 \cite{Meyer1999,Eisert1999}, have offered motivation and additional insight because quite often quantum strategies seem to achieve better results than classical ones \cite{Andronikos2018, Andronikos2021, Andronikos2022a}. The famous Prisoners' Dilemma game provides such the most prominent example \cite{Eisert1999,Giannakis2019}, which also applies to other abstract quantum games \cite{Giannakis2015a}. The quantization of many classical systems can even apply to political structures, as was shown in \cite{Andronikos2022}. To be fair, the quantum setting is not the only unconventional venue that has been pursued. The biological realm has also received its share of attention (see \cite{Theocharopoulou2019,Kastampolidou2020a,Kostadimas2021}). By drawing inspiration
from the biological domain, it is often possible to outperform classical strategies in games such as the Prisoners' Dilemma  \cite{Kastampolidou2020,Kastampolidou2021,Kastampolidou2021a,Papalitsas2021,Adam2023}.

To motivate the forthcoming presentation of the protocol we give a real-life example where the ability to have such an efficient and secure three-party information exchange protocol, complying with the specifications previously outline, is beneficial or even necessary.

\begin{example} [The undercover agents] \label{xmp:The Undercover Agents}
	This example should be seen as a proof of concept for more intricate real-life situations. Let us suppose that Bob and Charlie are two undercover agents working for a law enforcement agency. Information is critical for the success of their mission. Bob is desperately in need of some piece of information that knows Charlie has. He also knows that in their line of work the only real currency is information. So, he is ready to offer Charlie another piece of information that he believes will prove useful to Charlie. Bob and Charlie are not friends, just colleagues, and they do not fully trust each other, but they both trust their boss, Alice. Alice would be the ideal intermediary because she trusted by both agents, so she should be involved in the process of information exchange. Moreover, based on her involvement, Alice would be able to remember this secret transaction and vouch for her agents in case one, or both, of them would need in the future to prove that this particular information exchange took place, e.g., if their missions failed. At the end of the exchange process, Alice, should not have become aware of the contents of the information that was exchanged because both agents want to protect their informants. Finally, Alice stays at the headquarters, whereas Bob and Charlie work undercover in different parts of town, so it would be safer and more prudent if they did not meet in person at all. How can they complete this task?
	\hfill $\triangleleft$
\end{example}

\textbf{Contribution}. This paper presents a quantum protocol that solves efficiently and securely the problem of reciprocal information exchange between two information brokers. Its novelty is twofold: (i) in the integral use entanglement, and (ii) in the inclusion of a third party to supervise and facilitate the exchange process. Entanglement, one of the hallmarks of the quantum world, allows the simultaneous transmission of information by the two players, enhances the security of the operation, and also enables the spatially distributed execution of the protocol. There many practical situations, as briefly sketched in the previous example, where the presence of an intermediary would be required. The important observation here is that a trusted intermediary does not compromise the security of the procedure, and, after the completion of the protocol, remains oblivious of the actual information that was exchanged between the two players, i.e., there is no information leak whatsoever.

\subsection*{Organization} \label{subsec:Organization}

The structure of this paper is the following. Section \ref{sec:Introduction} contains an introduction to the subject and contains related references. Section \ref{sec:Preliminaries} gives a succinct reminder on entangled GHZ states. Section \ref{sec:The ESR IEP2 Protocol} provides a formal and detailed exposition of the quantum protocol, and, finally, Section \ref{sec:Discussion and Conclusions} contains a summary and a brief discussion on some of the finer points of this protocol.

\section{Preliminaries} \label{sec:Preliminaries}

Many quantum protocols can be described as game between well-known fictional players, commonly referred to as Alice, Bob, Charlie, who, while being spatially separated, are attempting to send and/or receive information securely. Typically, secure communication can be established via a combination of classical pairwise authenticated channels with pairwise quantum channels. Usually, the process of transmitting secret information takes place through the quantum channel using a multitude of different techniques, and, subsequently, in order to complete the process, a message is exchanged through a classical public channel. During this phase, an adversary who is mostly referred to as Eve, may appear and attempt to track this communication and steal any information possible. If such an eventuality, the major advantage of quantum cryptography over its classical counterpart is that during the transmission of information through the quantum channel, the communicating players are protected due to certain fundamental principles of quantum mechanics, such as the no-cloning theorem \cite{wootters1982single}, entanglement monogamy, etc.

Entanglement constitutes the fundamental basis of most quantum protocols, including the one proposed in this work, and possibly the de facto future of quantum cryptography, due to its numerous applications in the entire field. It is one of the fundamental principles of quantum mechanics and can be described mathematically as the linear combination of two or more product states. The Bell states are specific quantum states of two qubits, sometimes called an EPR pair, that represent the simplest examples of quantum entanglement. The entanglement of three or more qubits is referred to as a GHZ state. The fundamental idea of quantum entanglement in its simplest form is that it is possible for quantum particles to be entangled together and when a property is measured in one particle, it can be observed on the other particles instantaneously.

Contemporary quantum computers based on the circuit model can readily produce general states involving $n$ entangled qubits, which are denoted by $\ket{ GHZ_{n} }$. Implementing such a circuit requires $n$ qubits, one Hadamard gate that is applied to the first qubit, and $n - 1$ CNOT gates. We refer the interested reader to \cite{Cruz2019} for a practical methodology that can be utilized to construct efficient GHZ circuits, in the sense that it just takes $\lg n$ steps to produce the $\ket{ GHZ_{n} }$ state. For the proposed protocol $\ket{ GHZ_{3} }$ triplets suffice. A typical circuit capable of generating the $\ket{ GHZ_{3} }$ state is given in Figure \ref{fig:GHZ3_QC}. The circuit was designed using the IBM Quantum Composer \cite{IBMQuantumComposer2023}. The dotted lines are not part of the circuit; they just serve to provide a visual aid in order to distinguish ``time slices.'' Figure \ref{fig:GHZ3_SV}, also taken from the IBM Quantum Composer, depict the state vector description of the $\ket{ GHZ_{3} }$ state.

\begin{tcolorbox}
	[
		grow to left by = 0.00 cm,
		grow to right by = 0.00 cm,
		colback = WordTurquoiseLighter80!07,	
		enhanced jigsaw,						
		sharp corners,
		toprule = 1.0 pt,
		bottomrule = 1.0 pt,
		leftrule = 0.1 pt,
		rightrule = 0.1 pt,
		sharp corners,
		center title,
		fonttitle = \bfseries
	]
	\begin{figure}[H]
		\centering
		\includegraphics[scale = 0.75, trim = {0cm 0cm 0cm 0cm}, clip]{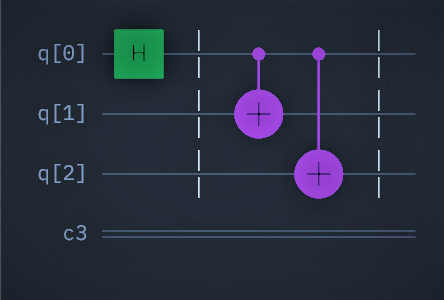}
		\caption{The above quantum circuit can entangle 3 qubits in the $\ket{ GHZ_3 } = \frac{ \ket{0} \ket{0} \ket{0} + \ket{1} \ket{1} \ket{1} } {\sqrt{2}}$ state. Any $\ket{ GHZ_n }$ state can be produced in an analogous manner.}
		\label{fig:GHZ3_QC}
	\end{figure}
\end{tcolorbox}

\begin{tcolorbox}
	[
		grow to left by = 0.25 cm,
		grow to right by = 0.25 cm,
		colback = white,			
		enhanced jigsaw,			
		sharp corners,
		toprule = 1.0 pt,
		bottomrule = 1.0 pt,
		leftrule = 0.1 pt,
		rightrule = 0.1 pt,
		sharp corners,
		center title,
		fonttitle = \bfseries
	]
	\centering
	\begin{figure}[H]
		\centering
		\includegraphics[scale = 0.50, trim = {0cm 4.5cm 2.0cm 0cm}, clip]{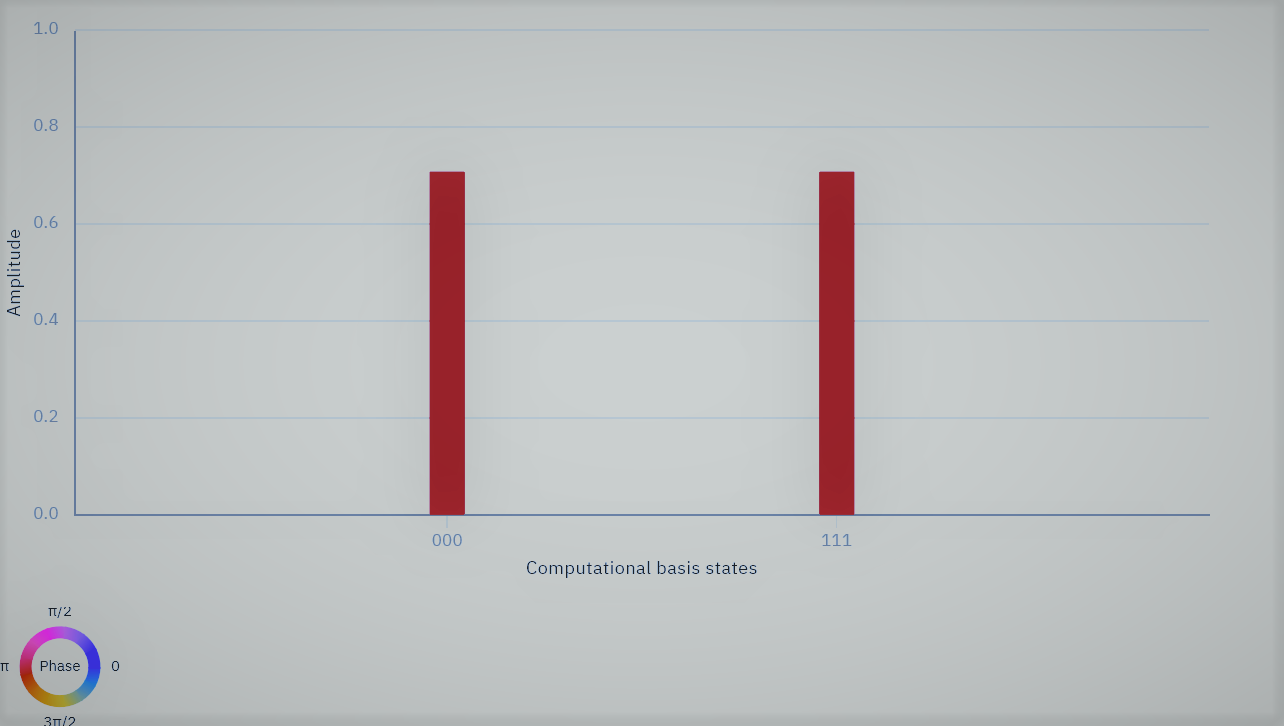}
		\caption{The above figure depicts the state vector description of the $\ket{ GHZ_3 }$ state.\\}
		\label{fig:GHZ3_SV}
	\end{figure}
\end{tcolorbox}

The mathematical description of the $\ket{ GHZ_{3} }$ state is given below.

\begin{align} \label{eq:GHZ_3 State}
	\ket{ GHZ_{3} }
	=
	\frac{ 1 }{ \sqrt{2} }
	\left( \ket{0}_{ A } \ket{0}_{ B } \ket{0}_{ C } + \ket{1}_{ A } \ket{1}_{ B } \ket{1}_{ C } \right)
	\ .
\end{align}

Our protocol requires not just a single $\ket{ GHZ_{3} }$ triplet, but $n$ such triplets. The state of a composite system comprised of $n$ $\ket{ GHZ_{3} }$ triplets is given by the next formula (for details we refer to \cite{Ampatzis2022} and \cite{Ampatzis2023}).

\begin{align} \label{eq:n GHZ_3 State}
	\ket{ GHZ_{3} }^{\otimes n}
	=
	\frac{1}{ \sqrt{2^n} }
	\sum_{\mathbf{x} \in \{ 0, 1 \}^n}
	\ket{\mathbf{x}}_{ A } \ket{\mathbf{x}}_{ B } \ket{\mathbf{x}}_{ C }
	\ .
\end{align}

In (\ref{eq:n GHZ_3 State}), ${\mathbf{x} \in \{ 0, 1 \}^n}$ ranges through all the $2^{n}$ basis kets, $\ket{\mathbf{x}}_{A}$, $\ket{\mathbf{x}}_{B}$ and $\ket{\mathbf{x}}_{C}$ correspond to the contents of Alice, Bob and Charlie's input registers, respectively.

\section{The protocol for simultaneous reciprocal information exchange between 2 players} \label{sec:The ESR IEP2 Protocol}

\begin{tcolorbox}
	[
		grow to left by = 0.00 cm,
		grow to right by = 0.00 cm,
		colback = MagentaVeryLight!20,		
		enhanced jigsaw,					
		sharp corners,
		toprule = 1.0 pt,
		bottomrule = 1.0 pt,
		leftrule = 0.1 pt,
		rightrule = 0.1 pt,
		sharp corners,
		center title,
		fonttitle = \bfseries
	]
	\begin{figure}[H]
		\centering
		\begin{tikzpicture} [scale = 1.00]
			\def \n {4}
			\def \Angle {360 / \n}
			\def \Radius {4}
			\def \Begin {3.8}
			\def \End {3.7}
			\node
			[
				shade, top color = WordAquaDarker25, bottom color = black, rectangle, text width = 10.00 cm, align = center
			] ( Label ) at ( 0.0, 7.50 )
			{ \color{white} \textbf{THE QUANTUM CHANNELS} \\
				Two quantum channels are required; one between Alice and Bob, and another between Alice and Charlie, where Alice, Bob and Charlie are pairwise spatially separated. Alice creates $n$ triplets in the $\ket{ GHZ_{3} }$ state. From each triplet, Alice keeps one qubit and sends through the quantum channels the other two to Bob and Charlie, one to each of them. };
			\node
			[
				charlie,
				scale = 2.00,
				anchor = center,
				label = { [ label distance = 0.00 cm ] east: \textbf{Charlie} }
			]
			( ) at ( { \Radius * cos(0 * \Angle) }, { \Radius * sin(0 * \Angle) } ) { };
			\node
			[
				alice,
				scale = 2.00,
				anchor = center,
				label = { [ label distance = 0.00 cm, text width = 5.00 cm, align = center ] north: \textbf{Alice\\(Source of the $\ket{GHZ_{3}}$ triples)} }
			]
			( ) at ( { \Radius * cos(1 * \Angle) }, { \Radius * sin(1 * \Angle) } ) { };
			
			\node
			[
				bob,
				scale = 2.00,
				anchor = center,
				label = { [ label distance = 0.00 cm ] west: \textbf{Bob} }
			]
			( ) at ( { \Radius * cos(2 * \Angle) }, { \Radius * sin(2 * \Angle) } ) { };
			\begin{scope}[on background layer]
				\foreach \angle in { 90 }
				\draw [ WordBlueVeryLight, ->, >=stealth, line width = 5.0 mm ] 
				( { \Begin * cos(\angle) }, { \Begin * sin(\angle) } ) --
				( { \End * cos(\angle + 90) }, { \End * sin(\angle + 90) } )
				node [ rotate = -45, midway, text = black, rotate = \angle ] {$n$ qubits};
				\draw [ WordBlueVeryLight, ->, >=stealth, line width = 5.0 mm ]
				( { \Begin * cos(90) }, { \Begin * sin(90) } ) --
				( { \End * cos(0) }, { \End * sin(0) } )
				node [ rotate = -45, midway, text = black ] {$n$ qubits};
			\end{scope}
		\end{tikzpicture}
		\caption{This figure shows Alice, Bob and Charlie, who are spatially separated, and the two quantum channels, one between Alice and Bob and the other between Alice and Charlie. Alice produces $n$ triplets of photons entangled in the $\ket{ GHZ_{3} }$ state. From each triplet Alice keeps one for herself, sends the second to Bob, and the third to Charlie through the corresponding quantum channels.} \label{fig:Alice, Bob and Charlie's Quantum Channels}
	\end{figure}
\end{tcolorbox}

In this section we present the entanglement-based protocol for simultaneous reciprocal information exchange between 2 players, or ESR for short. The ESR protocol is designed so as to enable two players, named Bob and Charlie, who are perceived as information brokers, to exchange information simultaneously and reciprocally. The process is mediated by Alice, who plays the mutually trusted intermediary. All our three protagonists are assumed to be spatially separated. The main idea is to achieve this task by using an appropriate number (which we denote by $n$) of maximally entangled $\ket{ GHZ_{3} }$ triplets. These are created by Alice, trusted by both Bob and Charlie, who evenly distributes all triplets among herself, Bob and Charlie, using two corresponding pairwise quantum channels. This results in each player having precisely one qubit from every triplet. Alice's role is crucial not only to the successful completion of the protocol, but also as a possible witness that the information exchange really took place. It is very important to point out that, despite being an integral part of the exchange, Alice does not end up knowing anything. The protocol can be conceptually divided in two parts, a quantum and a classical part, taking place through the quantum and the public classical channels, respectively.

\subsection{The quantum part of the protocol} \label{subsec:Quantum Part of the Protocol}

The game effectively begins after Alice, Bob and Charlie have populated their output registers, denoted by AOR, BOR and COR, respectively, with their $n$ qubits. The whole setting is depicted in Figure \ref{fig:Alice, Bob and Charlie's Quantum Channels}. The protocol itself can be implemented with the distributed quantum circuit of Figure \ref{fig:ESR IEP2 Protocol}. Although this circuit is distributed, since the players' local circuits are spatially separated, it is in fact one composite system because their input registers are strongly correlated due to the presence of entanglement. The notation employed in the circuit of Figure \ref{fig:ESR IEP2 Protocol} is explained in Table \ref{tbl:Figure ESR IEP2 QC Notations and Abbreviations}.

\begin{tcolorbox}
	[
		grow to left by = 1.50 cm,
		grow to right by = 1.50 cm,
		colback = MagentaVeryLight!12,			
		enhanced jigsaw,						
		sharp corners,
		toprule = 1.0 pt,
		bottomrule = 1.0 pt,
		leftrule = 0.1 pt,
		rightrule = 0.1 pt,
		sharp corners,
		center title,
		fonttitle = \bfseries
	]
	\begin{figure}[H]
		\centering
		\begin{tikzpicture}[ scale = 0.90 ]
			\begin{yquant}
				nobit AUX_C_0;
				qubits { $CIR$ } CIR;
				qubit { $COR$: \ $\ket{1}$ } COR;
				nobit AUX_C_1;
				nobit AUX_B_C;
				nobit AUX_B_0;
				qubits { $BIR$ } BIR;
				qubit { $BOR$: \ $\ket{1}$ } BOR;
				nobit AUX_B_1;
				nobit AUX_A_B;
				nobit AUX_A_0;
				qubits { $AIR$ } AIR;
				nobit AUX_A_1;
				nobit AUX_A_2;
				[ name = Ph0, WordBlueVeryLight, line width = 0.50 mm, label = Initial State ]
				barrier ( - ) ;
				[ draw = WordAquaLighter40, fill = WordAquaLighter40 ] [ radius = 0.5 cm ] box {\large\sf{H}} COR;
				[ draw = WordAquaLighter40, fill = WordAquaLighter40 ] [ radius = 0.5 cm ] box {\large\sf{H}} BOR;
				[ name = Ph1, WordBlueVeryLight, line width = 0.50 mm, label = Phase 1 ]
				barrier ( - ) ;
				[ draw = WordBlueDarker50, fill = WordBlueDarker50 ] [ x radius = 0.7 cm ] box {\color{white} {\large\sf{U}$_{f_{C}}$}} (CIR - COR);
				[ draw = WordBlueDarker50, fill = WordBlueDarker50 ] [ x radius = 0.7 cm ] box {\color{white} {\large\sf{U}$_{f_{B}}$}} (BIR - BOR);
				[ name = Ph2, WordBlueVeryLight, line width = 0.50 mm, label = Phase 2 ]
				barrier ( - ) ;
				[ draw = WordAquaLighter40, fill = WordAquaLighter40 ] [ radius = 0.5 cm ] box {\large\sf{H}$^{\otimes n}$} CIR;
				[ draw = WordAquaLighter40, fill = WordAquaLighter40 ] [ radius = 0.5 cm ] box {\large\sf{H}$^{\otimes n}$} BIR;
				[ draw = WordAquaLighter40, fill = WordAquaLighter40 ] [ radius = 0.5 cm ] box {\large\sf{H}$^{\otimes n}$} AIR;
				[ name = Ph3, WordBlueVeryLight, line width = 0.50 mm, label = Phase 3 ]
				barrier ( - ) ;
				[ fill = MagentaVeryLight!50 ] [ radius = 0.5 cm ] measure CIR;
				[ fill = MagentaVeryLight!50 ] [ radius = 0.5 cm ] measure BIR;
				[ fill = MagentaVeryLight!50 ] [ radius = 0.5 cm ] measure AIR;
				[ name = Ph4, WordBlueVeryLight, line width = 0.50 mm, label = Phase 4 ]
				barrier ( - ) ;
				hspace {0.3 cm} CIR;
				output {$\ket{ \mathbf{c} }$} CIR;
				output {$\ket{ \mathbf{b} }$} BIR;
				output {$\ket{ \mathbf{a} }$} AIR;
				\node [ below = 4.75 cm ] at (Ph0) {$\ket{\psi_{0}}$};
				\node [ below = 4.75 cm ] at (Ph1) {$\ket{\psi_{1}}$};
				\node [ below = 4.75 cm ] at (Ph2) {$\ket{\psi_{2}}$};
				\node [ below = 4.75 cm ] at (Ph3) {$\ket{\psi_{3}}$};
				\node [ below = 4.75 cm ] at (Ph4) {$\ket{\psi_{4}}$};
				\node
				[
					charlie,
					scale = 1.25,
					anchor = center,
					label = { [ label distance = 0.00 cm ] north: Charlie }
				]
				( ) at ( -2.75, -1.50 ) { };
				\node
				[
					bob,
					scale = 1.25,
					anchor = center,
					label = { [ label distance = 0.00 cm ] north: Bob }
				]
				( ) at ( -2.75, -5.10 ) { };
				\node
				[
					alice,
					scale = 1.25,
					anchor = center,
					label = { [ label distance = 0.00 cm ] north: Alice }
				]
				( ) at ( -2.75, -8.70 ) {\footnotesize Creates $n$ $\ket{ GHZ_{3} }$ triplets};
				
			\end{yquant}
			\begin{scope}[on background layer]
				\draw [MagentaDark, ->, >=stealth, line width = 1.2 mm] (-2.00, -8.80) -- (-2.00, -4.50) node [ above, rotate = 90, midway, text = black ] {$n$ qubits};
				\draw [MagentaDark, ->, >=stealth, line width = 1.25 mm] (-3.60, -8.80) -- (-3.60, -0.90) node [ above, rotate = 90, midway, text = black ] {$n$ qubits};
			\end{scope}
		\end{tikzpicture}
		\caption{This figure is an abstract visualization of the quantum circuits employed by Alice, Bob and Charlie. Although they are spatially separated, they are correlated, due to the phenomenon of entanglement. Thus, they form a composite system, whose temporal evolution is given by the state vectors $\ket{\psi_{0}}$, $\ket{\psi_{1}}$, $\ket{\psi_{2}}$, $\ket{\psi_{3}}$ and $\ket{\psi_{4}}$.}
		\label{fig:ESR IEP2 Protocol}
	\end{figure}
\end{tcolorbox}

\begin{tcolorbox}
	[
		grow to left by = 0.00 cm,
		grow to right by = 0.00 cm,
		colback = MagentaVeryLight!07,			
		enhanced jigsaw,						
		sharp corners,
		boxrule = 0.1 pt,
		toprule = 0.1 pt,
		bottomrule = 0.1 pt
	]
	\begin{table}[H]
		\renewcommand{\arraystretch}{1.40}
		\caption{This table contains the notations and abbreviations that are used in Figure \ref{fig:ESR IEP2 Protocol}.}
		\label{tbl:Figure ESR IEP2 QC Notations and Abbreviations}
		\centering
		\begin{tabular}
			{
				>{\centering\arraybackslash} m{2.50 cm} !{\vrule width 1.25 pt}
				>{\centering\arraybackslash} m{10.50 cm} 
			}
			\Xhline{7\arrayrulewidth}
			\multicolumn{2}{c}{\textbf{Notations and Abbreviations}}
			\\
			\Xhline{\arrayrulewidth}
			Symbolism
			&
			Explanation
			\\
			$n$
			&
			The number of qubits in each input register
			\\
			\Xhline{1\arrayrulewidth}
			AIR
			&
			Alice's $n$-qubit Input Register
			\\
			\Xhline{1\arrayrulewidth}
			BIR
			&
			Bob's $n$-qubit Input Register
			\\
			\Xhline{1\arrayrulewidth}
			BOR
			&
			Bob's single qubit Output Register
			\\
			\Xhline{1\arrayrulewidth}
			CIR
			&
			Charlie's $n$-qubit Input Register
			\\
			\Xhline{1\arrayrulewidth}
			COR
			&
			Charlie's single qubit Output Register
			\\
			\Xhline{7\arrayrulewidth}
		\end{tabular}
		\renewcommand{\arraystretch}{1.0}
	\end{table}
\end{tcolorbox}

The rigorous mathematical analysis of the ESR information exchange protocol invokes a couple of standard relations from the literature, typically found in most textbooks, like \cite{Nielsen2010} and \cite{Mermin2007}):

\begin{align}
	H \ket{1}
	&=
	\frac{1}{\sqrt{2}}
	\left( \ket{0} - \ket{1} \right)
	=
	\ket{-}
	\label{eq:Ket - Definition}
	\ , \quad \text{and}
	\\
	H^{\otimes n} \ket{ \mathbf{x} }
	&=
	\frac{1}{\sqrt{2^n}}
	\sum_{ \mathbf{z} \in \{ 0, 1 \}^n }
	(-1)^{ \mathbf{z \cdot x} } \ket{ \mathbf{z} }
	\ .
	\label{eq:Hadamard n-Fold Ket x}
\end{align}

In equation (\ref{eq:Hadamard n-Fold Ket x}) we have adhered to the typical convention of writing the contents of quantum registers in boldface, e.g., $\ket{ \mathbf{x} } = \ket{ x_{ n - 1 } } \dots \ket{ x_{0} }$, for some $n \geq 1$. Moreover, the notation $\mathbf{z \cdot x}$ stands for the \emph{inner product modulo} $2$, which, assuming that $\ket{ \mathbf{z} } = \ket{ z_{ n - 1 } } \dots \ket{ z_{0} }$, is defined as

\begin{align} \label{eq:Inner Product Modulo $2$}
	\mathbf{z \cdot x} = z_{ n - 1 } x_{ n - 1 } \oplus \dots \oplus z_{0} x_{0}
	\ .
\end{align}

Let us assume that $\mathbf{i}_{B}$ is the bit vector that represents the information that Bob possesses and intends to exchange with Charlie, and, symmetrically, that $\mathbf{i}_{C}$ is the bit vector corresponding to the information that Charlie possesses and intends to exchange with Bob. We define the \emph{auxiliary information bit vectors} $\widetilde{ \mathbf{i}_{B} }$ and $\widetilde{ \mathbf{i}_{C} }$ as follows, where the notation $| \cdot |$ denotes the length, i.e., number of bits, of the enclosed bit vector:

\begin{align}
	\widetilde{ \mathbf{i}_{B} }
	&=
	\mathbf{i}_{B} \
	\underbrace{ 0 \ \cdots \ 0 }_{ | \mathbf{i}_{C} | \rm\ times }
	\label{eq:Bob's Auxiliary Information Vector}
	\ , \quad \text{and}
	\\
	\widetilde{ \mathbf{i}_{C} }
	&=
	\underbrace{ 0 \ \cdots \ 0 }_{ | \mathbf{i}_{B} | \rm\ times }
	\ \mathbf{i}_{C}
	\label{eq:Charlie's Auxiliary Information Vector}
	\ .
\end{align}

We define the \emph{aggregated information bit vector} $\mathbf{i}$ as the concatenation of $\mathbf{i}_{B}$ and $\mathbf{i}_{C}$:

\begin{align}
	\mathbf{i}
	=
	\mathbf{i}_{B} \ \mathbf{i}_{C}
	\label{eq:Aggregated Information Vector}
	\ ,
\end{align}

and we set

\begin{align}
	n
	=
	| \mathbf{i} |
	=
	| \widetilde{ \mathbf{i}_{B} } |
	=
	| \widetilde{ \mathbf{i}_{C} } |
	=
	| \mathbf{i}_{B} | + | \mathbf{i}_{C} |
	\label{eq:Aggregated Information Vector Length}
	\ .
\end{align}

We use the boldface $ \mathbf{0}$ to abbreviate the zero bit vector of length $n$. Of course, we assume that $n$ is common knowledge to all three players. In effect, this can be easily achieved if Bob and Charlie share though the public channel the lengths $| \mathbf{i}_{B} |$ and $| \mathbf{i}_{C} |$ of their respective information bit vectors. This poses no danger whatsoever, as sharing the length does not reveal the contents of the secret information. The preceding discussion also implies that

\begin{align} \label{eq:Aggregated Information Vector Length as Mod $2$ Sum}
	\mathbf{i}
	=
	\widetilde{ \mathbf{i}_{B} }
	\oplus
	\widetilde{ \mathbf{i}_{C} }
	\ .
\end{align}

The initial state of the circuit of Figure \ref{fig:ESR IEP2 Protocol} is denoted by $\ket{ \psi_0 }$. In view of (\ref{eq:n GHZ_3 State}), $\ket{ \psi_0 }$ is given by

\begin{align} \label{eq:ESR IEP2 Phase 0}
	\ket{ \psi_0 }
	=
	\frac{1}{ \sqrt{2^n} }
	\sum_{ \mathbf{x} \in \{ 0, 1 \}^n }
	\ket{\mathbf{x}}_{ A }
	\ket{1}_{ B }
	\ket{\mathbf{x}}_{ B }
	\ket{1}_{ C }
	\ket{\mathbf{x}}_{ C }
	\ .
\end{align}

As is the trend nowadays, we stick to the Qiskit \cite{Qiskit2023} way of ordering the qubits, in which the most significant is the bottom qubit and the least significant the top. Furthermore, we rely on the subscripts $A$, $B$ and $C$ in order to designate Alice, Bob and Charlie's registers, respectively.

At the end of the first phase, Bob and Charlie have applied the Hadamard transform to their output registers and the resulting state has become $\ket{ \psi_1 }$:

\begin{align} \label{eq:ESR IEP2 Phase 1}
	\ket{\psi_1}
	\overset{(\ref{eq:Ket - Definition})}{=}
	\frac{1}{ \sqrt{2^n} }
	\sum_{ \mathbf{x} \in \{ 0, 1 \}^n }
	\ket{\mathbf{x}}_{ A }
	\ket{-}_{ B }
	\ket{\mathbf{x}}_{ B }
	\ket{-}_{ C }
	\ket{\mathbf{x}}_{ C }
	\ .
\end{align}

It is during the second phase that Bob and Charlie simultaneously encode the information they intend to exchange within the state of the quantum circuit. They do so by acting with their unitary transforms $U_{ f_{ B } }$ and $U_{ f_{ C } }$ on both their output and input registers. The unitary transforms $U_{ f_{ A } }$ and $U_{ f_{ B } }$ are constructed using the functions $f_{ B } ( \mathbf{x} )$ and $f_{ C } ( \mathbf{x} )$, respectively, according to the usual scheme

\begin{align}
	U_{ f_{ B } } &:
	\ket{ y }_{ B } \ket{ \mathbf{x} }_{ B }
	\rightarrow
	\ket{ y \oplus f_{ B }( \mathbf{x} ) }_{ B } \ket{ \mathbf{x} }_{ B }
	\label{eq:Bob's Transform}
	\ , \quad \text{and}
	\\
	U_{ f_{ C } } &:
	\ket{ y }_{ C } \ket{ \mathbf{x} }_{ C }
	\rightarrow
	\ket{ y \oplus f_{ C }( \mathbf{x} ) }_{ C } \ket{ \mathbf{x} }_{ C }
	\ .
	\label{eq:Charlie's Transform}
\end{align}

The functions  $f_{ B } ( \mathbf{x} )$ and $f_{ C } ( \mathbf{x} )$ are quite straightforward, relying on Bob and Charlie's auxiliary information bit vectors $\widetilde{ \mathbf{i}_{B} }$ and $\widetilde{ \mathbf{i}_{C} }$, respectively, according to the formulas (\ref{eq:Bob's Function}) and (\ref{eq:Charlie's Function}) presented below.

\begin{align}
	f_{ B } ( \mathbf{x} )
	&=
	\widetilde{ \mathbf{i}_{B} } \cdot \mathbf{x}
	\label{eq:Bob's Function}
	\ , \quad \text{and}
	\\
	f_{ C } ( \mathbf{x} )
	&=
	\widetilde{ \mathbf{i}_{C} } \cdot \mathbf{x}
	\ .
	\label{eq:Charlie's Function}
\end{align}

Therefore, (\ref{eq:Bob's Transform}) and (\ref{eq:Charlie's Transform}) can be explicitly written as

\begin{align}
	U_{ f_{ B } } &:
	\ket{ - }_{ B } \ket{ \mathbf{x} }_{ B }
	\rightarrow
	( - 1 )^{ \widetilde{ \mathbf{i}_{B} } \cdot \mathbf{x} }
	\ket{ - }_{ B } \ket{ \mathbf{x} }_{ B }
	\label{eq:Explicit Bob's Transform}
	\ , \quad \text{and}
	\\
	U_{ f_{ C } } &:
	\ket{ - }_{ C } \ket{ \mathbf{x} }_{ C }
	\rightarrow
	( - 1 )^{ \widetilde{ \mathbf{i}_{C} } \cdot \mathbf{x} }
	\ket{ - }_{ C } \ket{ \mathbf{x} }_{ C }
	\ .
	\label{eq:Explicit Charlie's Transform}
\end{align}

In view of the above calculations, at the end of the second phase the state of the quantum circuit has become $\ket{ \psi_2 }$:

\begin{align} \label{eq:ESR IEP2 Phase 2}
	\ket{\psi_2}
	&\ =
	\frac{ 1 }{ \sqrt{ 2^n } }
	\sum_{ \mathbf{x} \in \{ 0, 1 \}^n }
	\ket{ \mathbf{x} }_{ A }
	( - 1 )^{ \widetilde{ \mathbf{i}_{B} } \cdot \mathbf{x} }
	\ket{ - }_{ B } \ket{ \mathbf{x} }_{ B }
	( - 1 )^{ \widetilde{ \mathbf{i}_{C} } \cdot \mathbf{x} }
	\ket{ - }_{ C } \ket{ \mathbf{x} }_{ C }
	\nonumber \\
	&\ =
	\frac{ 1 }{ \sqrt{ 2^n } }
	\sum_{ \mathbf{x} \in \{ 0, 1 \}^n }
	( - 1 )^{ ( \widetilde{ \mathbf{i}_{B} } \oplus \widetilde{ \mathbf{i}_{C} } ) \cdot \mathbf{x} }
	\ket{ \mathbf{x} }_{ A }
	\ket{ - }_{ B } \ket{ \mathbf{x} }_{ B }
	\ket{ - }_{ C } \ket{ \mathbf{x} }_{ C }
	\nonumber \\
	&\overset{(\ref{eq:Aggregated Information Vector Length as Mod $2$ Sum})}{=}
	\frac{ 1 }{ \sqrt{ 2^n } }
	\sum_{ \mathbf{x} \in \{ 0, 1 \}^n }
	( - 1 )^{ \mathbf{i} \cdot \mathbf{x} }
	\ket{ \mathbf{x} }_{ A }
	\ket{ - }_{ B } \ket{ \mathbf{x} }_{ B }
	\ket{ - }_{ C } \ket{ \mathbf{x} }_{ C }
	\ .
\end{align}

As the third phase unfolds, Alice, Bob and Charlie apply their $n$-fold Hadamard transform to their input registers, driving the quantum circuit to the next state $\ket{\psi_3}$.

\begin{align} \label{eq:ESR IEP2 Phase 3 - I}
	\ket{\psi_3}
	&=
	\frac{ 1 }{ \sqrt{ 2^n } }
	\sum_{ \mathbf{x} \in \{ 0, 1 \}^n }
	( - 1 )^{ \mathbf{i} \cdot \mathbf{x} }
	H^{ \otimes n } \ket{ \mathbf{x} }_{A}
	\ket{ - }_{ B }
	H^{ \otimes n } \ket{ \mathbf{x} }_{B}
	\ket{ - }_{ C }
	H^{ \otimes n } \ket{ \mathbf{x} }_{C}
	\nonumber \\
	&\overset{(\ref{eq:Hadamard n-Fold Ket x})}{=}
	\frac{ 1 }{ \sqrt{ 2^n } }
	\sum_{ \mathbf{x} \in \{ 0, 1 \}^n }
	( - 1 )^{ \mathbf{i} \cdot \mathbf{x} }
	\left(
	\frac{ 1 }{ \sqrt{ 2^n } }
	\sum_{ \mathbf{a} \in \{ 0, 1 \}^n }
	( - 1 )^{ \mathbf{a} \cdot \mathbf{x} }
	\ket{ \mathbf{a} }_{A}
	\right)
	\nonumber \\
	&\hspace{2.125 cm}
	\ket{ - }_{ B }
	\left(
	\frac{ 1 }{ \sqrt{ 2^n } }
	\sum_{ \mathbf{b} \in \{ 0, 1 \}^n }
	( - 1 )^{ \mathbf{b} \cdot \mathbf{x} }
	\ket{ \mathbf{b} }_{ B }
	\right)
	\nonumber \\
	&\hspace{2.14 cm}
	\ket{ - }_{ C }
	\left(
	\frac{ 1 }{ \sqrt{ 2^n } }
	\sum_{ \mathbf{c} \in \{ 0, 1 \}^n }
	( - 1 )^{ \mathbf{c} \cdot \mathbf{x} }
	\ket{ \mathbf{c} }_{ C }
	\right)
	\nonumber \\
	&=
	\frac{ 1 }{ 2^{ 2 n } }
	\sum_{ \mathbf{x} \in \{ 0, 1 \}^n }
	\sum_{ \mathbf{a} \in \{ 0, 1 \}^n }
	\sum_{ \mathbf{b} \in \{ 0, 1 \}^n }
	\sum_{ \mathbf{c} \in \{ 0, 1 \}^n }
	( - 1 )^{ ( \mathbf{i} \oplus \mathbf{a} \oplus \mathbf{b} \oplus \mathbf{c} ) \cdot \mathbf{x} }
	\ket{ \mathbf{a} }_{ A }
	\ket{ - }_{ B } \ket{ \mathbf{b} }_{ B }
	\ket{ - }_{ C } \ket{ \mathbf{b} }_{ C }
	\nonumber \\
	&=
	\frac{ 1 }{ 2^{ 2 n } }
	\sum_{ \mathbf{a} \in \{ 0, 1 \}^n }
	\sum_{ \mathbf{b} \in \{ 0, 1 \}^n }
	\sum_{ \mathbf{c} \in \{ 0, 1 \}^n }
	\sum_{ \mathbf{x} \in \{ 0, 1 \}^n }
	( - 1 )^{ ( \mathbf{i} \oplus \mathbf{a} \oplus \mathbf{b} \oplus \mathbf{c} ) \cdot \mathbf{x} }
	\ket{ \mathbf{a} }_{ A }
	\ket{ - }_{ B } \ket{ \mathbf{b} }_{ B }
	\ket{ - }_{ C } \ket{ \mathbf{b} }_{ C }
	\ .
\end{align}

We can express (\ref{eq:ESR IEP2 Phase 3 - I}) in a more intelligible form by invoking a useful property of the inner product modulo $2$. Whenever $\mathbf{c}$ is a fixed element of $\{ 0, 1 \}^n$, but different from $\mathbf{0}$, then for half of the elements $\mathbf{x} \in \{ 0, 1 \}^n$, \ $\mathbf{c} \cdot \mathbf{x}$ is $0$ and for the other half, \ $\mathbf{c} \cdot \mathbf{x}$ is $1$. However, when $\mathbf{c} = \mathbf{0}$, then for all $\mathbf{x} \in \{ 0, 1 \}^n$, $\ \mathbf{c} \cdot \mathbf{x} = 0$. For a more detailed analysis on this point we refer to \cite{Ampatzis2023}. Therefore, if

\begin{align} \label{eq:Fundamental Correlation Property}
	\mathbf{a} \oplus \mathbf{b} \oplus \mathbf{c} = \mathbf{i} \ ,
\end{align}

the sum $\sum_{ \mathbf{x} \in \{ 0, 1 \}^n } ( - 1 )^{ ( \mathbf{i} \oplus \mathbf{a} \oplus \mathbf{b} \oplus \mathbf{c} ) \cdot \mathbf{x} } \ket{ \mathbf{a} }_{ A } \ket{ - }_{ B } \ket{ \mathbf{b} }_{ B } \ket{ - }_{ C } \ket{ \mathbf{b} }_{ C }$ is equal to $2^{n}$ $\ket{ \mathbf{a} }_{ A }$ $\ket{ - }_{ B } \ket{ \mathbf{b} }_{ B }$ $\ket{ - }_{ C } \ket{ \mathbf{b} }_{ C }$, whereas if $\mathbf{a} \oplus \mathbf{b} \oplus \mathbf{c} \neq \mathbf{i}$, the sum reduces to $0$. Ergo, $\ket{\psi_3}$ can be written simpler as

\begin{align} \label{eq:ESR IEP2 Phase 3 - II}
	\hspace{-1 cm}
	\ket{\psi_3}
	=
	\frac{ 1 }{ 2^{ n } }
	\sum_{ \mathbf{a} \in \{ 0, 1 \}^n }
	\sum_{ \mathbf{b} \in \{ 0, 1 \}^n }
	\sum_{ \mathbf{c} \in \{ 0, 1 \}^n }
	\ket{ \mathbf{a} }_{ A }
	\ket{ - }_{ B } \ket{ \mathbf{b} }_{ B }
	\ket{ - }_{ C } \ket{ \mathbf{b} }_{ C }
	\ , \quad \text{where} \quad
	\mathbf{a} \oplus \mathbf{b} \oplus \mathbf{c} = \mathbf{i}
	\ .
\end{align}

We call relation (\ref{eq:Fundamental Correlation Property}) the \textbf{Fundamental Correlation Property} that intertwines the contents of Alice, Bob and Charlie's input registers. The intuition behind this property is that the entanglement between their input registers in the initialization of the quantum circuit, has caused this specific dependency that disallows the contents of the input registers from varying independently of its other.

The final measurement of their input registers by Alice, Bob and Charlie, drives the quantum circuit into its final state $\ket{\psi_4}$.

\begin{align}
	\label{eq:ESR IEP2 Final Measurement}
	\ket{\psi_4}
	=
	\ket{ \mathbf{a} }_{ A }
	\ket{ - }_{ B } \ket{ \mathbf{b} }_{ B }
	\ket{ - }_{ C } \ket{ \mathbf{b} }_{ C }
	\ , \quad \text{for some} \quad
	\mathbf{a}, \mathbf{b}, \mathbf{c} \in \{ 0, 1 \}^n
	\ ,
\end{align}

where $\mathbf{a}, \mathbf{b}$ and $\mathbf{c}$ are correlated via (\ref{eq:Fundamental Correlation Property}). One may regard the final contents $\mathbf{a}$ and $\mathbf{b}$ of Alice and Bob's input registers random, but in that case the final contents $\mathbf{c}$ of Charlie's input register are completely determined. Symmetrically, one may regard the final contents $\mathbf{b}$ and $\mathbf{c}$ of Bob and Charlie's input register random, in which case the final contents $\mathbf{a}$ of Alice's input register are completely determined, and so on.

\subsection{The classical part of the protocol} \label{subsec:Classical Part of the Protocol}

\begin{tcolorbox}
	[
		grow to left by = 0.00 cm,
		grow to right by = 0.00 cm,
		colback = WordAquaLighter80!50,		
		enhanced jigsaw,					
		sharp corners,
		toprule = 1.0 pt,
		bottomrule = 1.0 pt,
		leftrule = 0.1 pt,
		rightrule = 0.1 pt,
		sharp corners,
		center title,
		fonttitle = \bfseries
	]
	\begin{figure}[H]
		\centering
		\begin{tikzpicture} [scale = 1.00]
			\def \n {4}
			\def \Angle {360 / \n}
			\def \Radius {4}
			\def \Begin {3.8}
			\def \End {3.5}
			\node
			[
				shade, top color = GreenTeal, bottom color = black, rectangle, text width = 10.00 cm, align = center
			] ( Label ) at ( 0.0, 7.50 )
			{ \color{white} \textbf{THE CLASSICAL CHANNELS} \\
				Alice uses two classical channels, the first between herself and Bob, and the second between herself and Charlie, to send them the contents of her input register. Bob and Charlie use a third classical channel to exchange information. };
			\node
			[
				charlie,
				scale = 2.00,
				anchor = center,
				label = { [ label distance = 0.00 cm ] east: \textbf{Charlie} }
			]
			( ) at ( { \Radius * cos(0 * \Angle) }, { \Radius * sin(0 * \Angle) } ) { };
			\node
			[
				alice,
				scale = 2.00,
				anchor = center,
				label = { [ label distance = 0.00 cm, text width = 5.00 cm, align = center ] north: \textbf{Alice\\(Source of the $\ket{GHZ_{3}}$ triples)} }
			]
			( ) at ( { \Radius * cos(1 * \Angle) }, { \Radius * sin(1 * \Angle) } ) { };
			
			\node
			[
				bob,
				scale = 2.00,
				anchor = center,
				label = { [ label distance = 0.00 cm ] west: \textbf{Bob} }
			]
			( ) at ( { \Radius * cos(2 * \Angle) }, { \Radius * sin(2 * \Angle) } ) { };
			\begin{scope}[on background layer]
				\foreach \angle in { 90 }
				\draw [ Purple, ->, >=stealth, line width = 3.5 mm ] 
				( { \Radius * cos(\angle) }, { \Radius * sin(\angle) } ) --
				( { \End * cos(\angle + 85) }, { \End * sin(\angle + 85) } )
				node [ rotate = 45, midway, text = white ] {$\mathbf{a}_{C}$};
				\draw [ Purple, ->, >=stealth, line width = 3.5 mm ]
				( { \Radius * cos(90) }, { \Radius * sin(90) } ) --
				( { \End * cos(5) }, { \End * sin(5) } )
				node [ rotate = -45, midway, text = white ] {$\mathbf{a}_{B}$};
				\draw [ MagentaDark, ->, >=stealth, line width = 3.5 mm ] 
				( { \End * cos(180) }, { \End * sin(180) + 0.1 } ) --
				( { \End * cos(0) }, { \End * sin(0) + 0.1 } )
				node [ midway, text = white ] {$\mathbf{b}_{B}$};
				\draw [ MagentaDark, ->, >=stealth, line width = 3.5 mm ] 
				( { \End * cos(0) }, { \End * sin(0) - 0.5 } ) --
				( { \End * cos(180) }, { \End * sin(180) - 0.5 } )
				node [ midway, text = white ] {$\mathbf{c}_{C}$};
			\end{scope}
		\end{tikzpicture}
		\caption{This figure shows Alice, Bob and Charlie, who are spatially separated, and their three classical channels, the first between Alice and Bob, the second between Alice and Charlie, and the third between Bob and Charlie.} \label{fig:Alice, Bob and Charlie's Classical Channels}
	\end{figure}
\end{tcolorbox}

To complete the ESR information exchange protocol one final step remains, and this step takes place in the classical public channels. We may write the contents of the players' input registers as follows:

\begin{align}
	\mathbf{a}
	&=
	\mathbf{a}_{B} \ \mathbf{a}_{C},
	&\text{where} \
	| \mathbf{a}_{B} | = | \mathbf{i}_{B} | \
	&\text{and} \
	| \mathbf{a}_{C} | = | \mathbf{i}_{C} |
	\label{eq:Alice's Input Register Final Contents}
	\\
	\mathbf{b}
	&=
	\mathbf{b}_{B} \ \mathbf{b}_{C},
	& \text{where} \
	| \mathbf{b}_{B} | = | \mathbf{i}_{B} | \
	&\text{and} \
	| \mathbf{b}_{C} | = | \mathbf{i}_{C} |
	\label{eq:Bob's Input Register Final Contents}
	\ , \quad \text{and}
	\\
	\mathbf{c}
	&=
	\mathbf{c}_{B} \ \mathbf{c}_{C},
	&\text{where} \
	| \mathbf{c}_{B} | = | \mathbf{i}_{B} | \
	&\text{and} \
	| \mathbf{c}_{C} | = | \mathbf{i}_{C} |
	\label{eq:Charlie's Input Register Final Contents}
	\ .
\end{align}

The previous formulas, allow us to refine equation (\ref{eq:Fundamental Correlation Property}) into two independent parts, the first regarding the information bit vector $\mathbf{i}_{B}$ that Bob intends to communicate to Charlie, and the second regarding the information bit vector $\mathbf{i}_{C}$, that Charlie intends to communicate to Bob.

\begin{align}
	\mathbf{a}_{B} \oplus \mathbf{b}_{B} \oplus \mathbf{c}_{B}
	&=
	\mathbf{i}_{B}
	\label{eq:Bob's Information Fundamental Correlation Property}
	\ , \quad \text{and}
	\\
	\mathbf{a}_{C} \oplus \mathbf{b}_{C} \oplus \mathbf{c}_{C}
	&=
	\mathbf{i}_{C}
	\label{eq:Charlie's Information Fundamental Correlation Property}
	\ .
\end{align}

The above relations (\ref{eq:Bob's Information Fundamental Correlation Property}) and (\ref{eq:Charlie's Information Fundamental Correlation Property}) dictate what remains to be done in the final classical part of the ESR information exchange protocol so that Bob and Charlie receive the intended information.

\begin{itemize}
	\item	Alice must use two classical public channels to communicate with Bob and Charlie. Specifically, she must send through these channels the information bit vectors $\mathbf{a}_{C}$ and $\mathbf{a}_{B}$ to Bob and Charlie, respectively.
	\item	Bob and Charlie must use a third classical channel to communicate with each other. This communication must take place with caution. They must not reveal the entire contents of their input registers because then Alice, and any other adversary for that matter, will be able to piece together the secret information $\mathbf{i}_{B}$ and $\mathbf{i}_{C}$. They must transmit only the absolutely necessary information for the successful completion of the exchange. This means that Bob must send to Charlie only the information vector $\mathbf{b}_{B}$ and not the whole contents $\mathbf{b}$ of his input register. Reciprocally, Charlie must send to Bob only the information vector $\mathbf{c}_{C}$ and not the whole contents $\mathbf{c}$ of his input register.
\end{itemize}

After the communications outlined above have taken place, the ESR protocol is concluded. Bob knows $\mathbf{a}_{C}$, $\mathbf{c}_{C}$, and, of course the contents of his input register, and can discover the secret information $\mathbf{i}_{C}$, according to equation (\ref{eq:Charlie's Information Fundamental Correlation Property}). Symmetrically, Charlie, being in possession of $\mathbf{a}_{B}$, $\mathbf{b}_{B}$, can reconstruct $\mathbf{i}_{B}$, as equation (\ref{eq:Bob's Information Fundamental Correlation Property}) asserts. Alice, despite her critical contribution in the implementation of the protocol, lacks the necessary information, namely $\mathbf{b}_{C}$ and $\mathbf{c}_{B}$, which has not been made public. Thus, she is no position to uncover either $\mathbf{i}_{B}$, or $\mathbf{i}_{C}$.

\section{Discussion and conclusions} \label{sec:Discussion and Conclusions}

This work proposed the ESR protocol for the simultaneous reciprocal exchange of secret information between Bob and Charlie. The whole process is mediated by Alice, who is assumed to be a trusted intermediary. Alice is necessary for the completion of the ESR protocol because without the contents of her input register, the information can be reconstructed. Although her contribution is absolutely essential, proper execution of the last classical part of the protocol ensures that she gains no insight whatsoever about the information that was exchanged. Moreover, since Alice is assumed to be a player that is mutually trusted by both Bob and Charlie, her involvement does not compromise the security of the protocol.

\begin{tcolorbox}
	[
		grow to left by = 1.50 cm,
		grow to right by = 1.50 cm,
		colback = WordLightGreen!07,			
		enhanced jigsaw,						
		sharp corners,
		toprule = 1.0 pt,
		bottomrule = 1.0 pt,
		leftrule = 0.1 pt,
		rightrule = 0.1 pt,
		sharp corners,
		center title,
		fonttitle = \bfseries
	]
	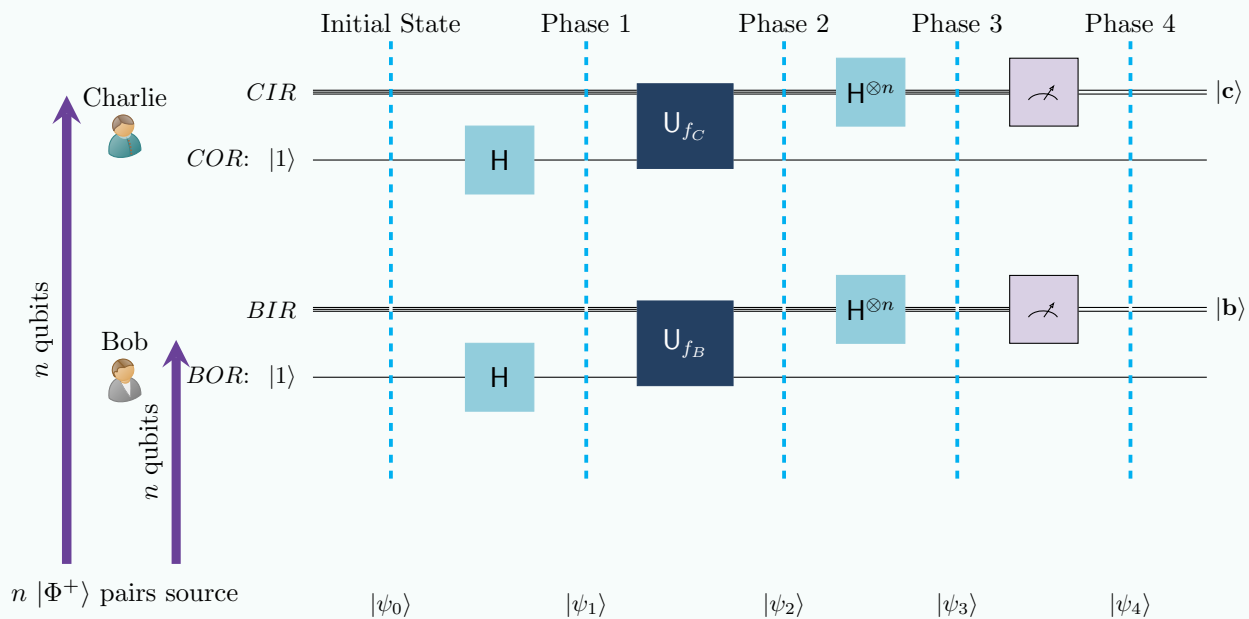
\begin{figure}[H]
		\centering
		\begin{tikzpicture}[ scale = 0.90 ]
			\begin{yquant}
				nobit AUX_C_0;
				qubits { $CIR$ } CIR;
				qubit { $COR$: \ $\ket{1}$ } COR;
				nobit AUX_C_1;
				nobit AUX_B_C;
				nobit AUX_B_0;
				qubits { $BIR$ } BIR;
				qubit { $BOR$: \ $\ket{1}$ } BOR;
				nobit AUX_B_1;
				nobit AUX_A_B;
				nobit AUX_A_2;
				[ name = Ph0, WordBlueVeryLight, line width = 0.50 mm, label = Initial State ]
				barrier ( - ) ;
				[ draw = WordAquaLighter40, fill = WordAquaLighter40 ] [ radius = 0.5 cm ] box {\large\sf{H}} COR;
				[ draw = WordAquaLighter40, fill = WordAquaLighter40 ] [ radius = 0.5 cm ] box {\large\sf{H}} BOR;
				[ name = Ph1, WordBlueVeryLight, line width = 0.50 mm, label = Phase 1 ]
				barrier ( - ) ;
				[ draw = WordBlueDarker50, fill = WordBlueDarker50 ] [ x radius = 0.7 cm ] box {\color{white} {\large\sf{U}$_{f_{C}}$}} (CIR - COR);
				[ draw = WordBlueDarker50, fill = WordBlueDarker50 ] [ x radius = 0.7 cm ] box {\color{white} {\large\sf{U}$_{f_{B}}$}} (BIR - BOR);
				[ name = Ph2, WordBlueVeryLight, line width = 0.50 mm, label = Phase 2 ]
				barrier ( - ) ;
				[ draw = WordAquaLighter40, fill = WordAquaLighter40 ] [ radius = 0.5 cm ] box {\large\sf{H}$^{\otimes n}$} CIR;
				[ draw = WordAquaLighter40, fill = WordAquaLighter40 ] [ radius = 0.5 cm ] box {\large\sf{H}$^{\otimes n}$} BIR;
				[ name = Ph3, WordBlueVeryLight, line width = 0.50 mm, label = Phase 3 ]
				barrier ( - ) ;
				[ fill = MagentaVeryLight!50 ] [ radius = 0.5 cm ] measure CIR;
				[ fill = MagentaVeryLight!50 ] [ radius = 0.5 cm ] measure BIR;
				[ name = Ph4, WordBlueVeryLight, line width = 0.50 mm, label = Phase 4 ]
				barrier ( - ) ;
				hspace {0.3 cm} CIR;
				output {$\ket{ \mathbf{c} }$} CIR;
				output {$\ket{ \mathbf{b} }$} BIR;
				\node [ below = 4.75 cm ] at (Ph0) {$\ket{\psi_{0}}$};
				\node [ below = 4.75 cm ] at (Ph1) {$\ket{\psi_{1}}$};
				\node [ below = 4.75 cm ] at (Ph2) {$\ket{\psi_{2}}$};
				\node [ below = 4.75 cm ] at (Ph3) {$\ket{\psi_{3}}$};
				\node [ below = 4.75 cm ] at (Ph4) {$\ket{\psi_{4}}$};
				\node
				[
					charlie,
					scale = 1.25,
					anchor = center,
					label = { [ label distance = 0.00 cm ] north: Charlie }
				]
				( ) at ( -2.75, -1.50 ) { };
				\node
				[
					bob,
					scale = 1.25,
					anchor = center,
					label = { [ label distance = 0.00 cm ] north: Bob }
				]
				( ) at ( -2.75, -5.10 ) { };
				\node
				[
					anchor = center,
					label = {
						[ label distance = 0.00 cm ] north:$n$ $\ket{\Phi^{+}}$ pairs source
					}
				]
				( ) at ( -2.75, -8.70 ) { };
			\end{yquant}
			\begin{scope}[on background layer]
				\draw [MagentaDark, ->, >=stealth, line width = 1.2 mm] (-2.00, -7.80) -- (-2.00, -4.50) node [ above, rotate = 90, midway, text = black ] {$n$ qubits};
				\draw [MagentaDark, ->, >=stealth, line width = 1.25 mm] (-3.60, -7.80) -- (-3.60, -0.90) node [ above, rotate = 90, midway, text = black ] {$n$ qubits};
			\end{scope}
		\end{tikzpicture}
		\caption{This figure outlines a distributed quantum circuit, where only Bob and Charlie are present. Again both are correlated, due to the phenomenon of entanglement. Thus, they form a composite system, whose temporal evolution is given by the state vectors $\ket{\psi_{0}}$, $\ket{\psi_{1}}$, $\ket{\psi_{2}}$, $\ket{\psi_{3}}$ and $\ket{\psi_{4}}$.}
		\label{fig:ESR IEP2 Protocol w/o Alice}
	\end{figure}
\end{tcolorbox}

Of course, one can easily envision a protocol for information exchange between Bob and Charlie that does not involve Alice at all. The ESR protocol can be easily simplified to function without the presence of Alice, using for instance the quantum circuit shown in Figure \ref{fig:ESR IEP2 Protocol w/o Alice}. In such a case, instead of entangled triplets, it would be necessary to employ entangled pairs, e.g., EPR pairs in the Bell $\ket{\Phi^{+}}$ state:

\begin{align} \label{eq:Bell State Phi +}
	\ket{\Phi^{+}} = \frac{ \ket{0}_{B} \ket{0}_{C} + \ket{1}_{B} \ket{1}_{C} } {\sqrt{2}}
	\ .
\end{align}

Such an approach would certainly require a somewhat simpler quantum circuit for the production of the EPR pairs. Furthermore, the whole mathematical description of the protocol would be considerably easier. Nonetheless one would still have to address the requirement for a trusted source to produce and distribute that $n$ $\ket{\Phi^{+}}$ pairs to Bob and Charlie.

However, it is our belief that, as we have advocated in the Introduction, there are many real-life situations where having a third party witnessing and verifying that such a transaction has indeed occurred can be beneficial, or even necessary. Ergo, it seems prudent to able to account for this eventuality, particularly in view of the fact that the third party not only is trusted, but it does not gain knowledge of the secret information that was exchanged.

\bibliographystyle{ieeetr}
\bibliography{ESRIEP2}

\begin{thebibliography}{10}

\bibitem{chen2016report}
L.~Chen, L.~Chen, S.~Jordan, Y.-K. Liu, D.~Moody, R.~Peralta, R.~Perlner, and
  D.~Smith-Tone, {\em Report on post-quantum cryptography}, vol.~12.
\newblock US Department of Commerce, National Institute of Standards and
  Technology, 2016.

\bibitem{alagic2019status}
G.~Alagic, G.~Alagic, J.~Alperin-Sheriff, D.~Apon, D.~Cooper, Q.~Dang, Y.-K.
  Liu, C.~Miller, D.~Moody, R.~Peralta, {\em et~al.}, {\em Status report on the
  first round of the NIST post-quantum cryptography standardization process}.
\newblock US Department of Commerce, National Institute of Standards and
  Technology~…, 2019.

\bibitem{alagic2020status}
G.~Alagic, J.~Alperin-Sheriff, D.~Apon, D.~Cooper, Q.~Dang, J.~Kelsey, Y.-K.
  Liu, C.~Miller, D.~Moody, R.~Peralta, {\em et~al.}, ``Status report on the
  second round of the nist post-quantum cryptography standardization process,''
  {\em US Department of Commerce, NIST}, 2020.

\bibitem{alagic2022status}
G.~Alagic, D.~Apon, D.~Cooper, Q.~Dang, T.~Dang, J.~Kelsey, J.~Lichtinger,
  C.~Miller, D.~Moody, R.~Peralta, {\em et~al.}, ``Status report on the third
  round of the nist post-quantum cryptography standardization process,'' {\em
  National Institute of Standards and Technology, Gaithersburg}, 2022.

\bibitem{Shor1994}
P.~Shor, ``Algorithms for quantum computation: discrete logarithms and
  factoring,'' in {\em Proceedings 35th Annual Symposium on Foundations of
  Computer Science}, {IEEE} Comput. Soc. Press, 1994.

\bibitem{Grover1996}
L.~Grover, ``A fast quantum mechanical algorithm for database search,'' in {\em
  Proc. of the Twenty-Eighth Annual ACM Symposium on the Theory of Computing,
  1996}, 1996.

\bibitem{IBMEagle}
J.~Chow, O.~Dial, and J.~Gambetta, ``{IBM} {Quantum} breaks the 100-qubit
  processor barrier.''
  \url{https://research.ibm.com/blog/127-qubit-quantum-processor-eagle}, 2021.
\newblock Accessed: 2022-04-03.

\bibitem{IBMOsprey}
I.~Newsroom, ``{IBM} unveils 400 qubit-plus quantum processor.''
  \url{https://newsroom.ibm.com/2022-11-09-IBM-Unveils-400-Qubit-Plus-Quantum-Processor-and-Next-Generation-IBM-Quantum-System-Two},
  2022.
\newblock Accessed: 2022-04-03.

\bibitem{chamola2021information}
V.~Chamola, A.~Jolfaei, V.~Chanana, P.~Parashari, and V.~Hassija, ``Information
  security in the post quantum era for 5g and beyond networks: Threats to
  existing cryptography, and post-quantum cryptography,'' {\em Computer
  Communications}, vol.~176, pp.~99--118, 2021.

\bibitem{Ekert1991}
A.~K. Ekert, ``Quantum cryptography based on bell's theorem,'' {\em Physical
  Review Letters}, vol.~67, no.~6, pp.~661--663, 1991.

\bibitem{Bennett1992}
C.~H. Bennett, G.~Brassard, and N.~D. Mermin, ``Quantum cryptography without
  bell's theorem,'' {\em Physical Review Letters}, vol.~68, no.~5,
  pp.~557--559, 1992.

\bibitem{Gisin2004}
N.~Gisin, G.~Ribordy, H.~Zbinden, D.~Stucki, N.~Brunner, and V.~Scarani,
  ``Towards practical and fast quantum cryptography,'' {\em arXiv preprint
  quant-ph/0411022}, 2004.

\bibitem{inoue2002differential}
K.~Inoue, E.~Waks, and Y.~Yamamoto, ``Differential phase shift quantum key
  distribution,'' {\em Physical review letters}, vol.~89, no.~3, p.~037902,
  2002.

\bibitem{guan2015experimental}
J.-Y. Guan, Z.~Cao, Y.~Liu, G.-L. Shen-Tu, J.~S. Pelc, M.~Fejer, C.-Z. Peng,
  X.~Ma, Q.~Zhang, and J.-W. Pan, ``Experimental passive round-robin
  differential phase-shift quantum key distribution,'' {\em Physical review
  letters}, vol.~114, no.~18, p.~180502, 2015.

\bibitem{waks2006security}
E.~Waks, H.~Takesue, and Y.~Yamamoto, ``Security of differential-phase-shift
  quantum key distribution against individual attacks,'' {\em Physical Review
  A}, vol.~73, no.~1, p.~012344, 2006.

\bibitem{Ampatzis2021}
M.~Ampatzis and T.~Andronikos, ``{QKD} based on symmetric entangled
  bernstein-vazirani,'' {\em Entropy}, vol.~23, no.~7, p.~870, 2021.

\bibitem{Ampatzis2022}
M.~Ampatzis and T.~Andronikos, ``A symmetric extensible protocol for quantum
  secret sharing,'' {\em Symmetry}, vol.~14, no.~8, p.~1692, 2022.

\bibitem{attasena2017secret}
V.~Attasena, J.~Darmont, and N.~Harbi, ``Secret sharing for cloud data
  security: a survey,'' {\em The VLDB Journal}, vol.~26, no.~5, pp.~657--681,
  2017.

\bibitem{ermakova2013secret}
T.~Ermakova and B.~Fabian, ``Secret sharing for health data in multi-provider
  clouds,'' in {\em 2013 IEEE 15th conference on business informatics},
  pp.~93--100, IEEE, 2013.

\bibitem{cha2021blockchain}
J.~Cha, S.~K. Singh, T.~W. Kim, and J.~H. Park, ``Blockchain-empowered cloud
  architecture based on secret sharing for smart city,'' {\em Journal of
  Information Security and Applications}, vol.~57, p.~102686, 2021.

\bibitem{karlsson1999quantum}
A.~Karlsson, M.~Koashi, and N.~Imoto, ``Quantum entanglement for secret sharing
  and secret splitting,'' {\em Physical Review A}, vol.~59, no.~1, p.~162,
  1999.

\bibitem{smith2000quantum}
A.~D. Smith, ``Quantum secret sharing for general access structures,'' {\em
  arXiv preprint quant-ph/0001087}, 2000.

\bibitem{gottesman2000theory}
D.~Gottesman, ``Theory of quantum secret sharing,'' {\em Physical Review A},
  vol.~61, no.~4, p.~042311, 2000.

\bibitem{fortescue2012reducing}
B.~Fortescue and G.~Gour, ``Reducing the quantum communication cost of quantum
  secret sharing,'' {\em IEEE transactions on information theory}, vol.~58,
  no.~10, pp.~6659--6666, 2012.

\bibitem{qin2020hierarchical}
H.~Qin, W.~K. Tang, and R.~Tso, ``Hierarchical quantum secret sharing based on
  special high-dimensional entangled state,'' {\em IEEE Journal of Selected
  Topics in Quantum Electronics}, vol.~26, no.~3, pp.~1--6, 2020.

\bibitem{senthoor2022theory}
K.~Senthoor and P.~K. Sarvepalli, ``Theory of communication efficient quantum
  secret sharing,'' {\em IEEE Transactions on Information Theory}, 2022.

\bibitem{fu2015long}
Y.~Fu, H.-L. Yin, T.-Y. Chen, and Z.-B. Chen, ``Long-distance
  measurement-device-independent multiparty quantum communication,'' {\em
  Physical review letters}, vol.~114, no.~9, p.~090501, 2015.

\bibitem{wu2020passive}
X.~Wu, Y.~Wang, and D.~Huang, ``Passive continuous-variable quantum secret
  sharing using a thermal source,'' {\em Physical Review A}, vol.~101, no.~2,
  p.~022301, 2020.

\bibitem{grice2019quantum}
W.~P. Grice and B.~Qi, ``Quantum secret sharing using weak coherent states,''
  {\em Physical Review A}, vol.~100, no.~2, p.~022339, 2019.

\bibitem{gu2021secure}
J.~Gu, Y.-M. Xie, W.-B. Liu, Y.~Fu, H.-L. Yin, and Z.-B. Chen, ``Secure quantum
  secret sharing without signal disturbance monitoring,'' {\em Optics Express},
  vol.~29, no.~20, pp.~32244--32255, 2021.

\bibitem{Bennett1984}
C.~H. Bennett and G.~Brassard, ``Quantum cryptography: Public key distribution
  and coin tossing,'' in {\em Proceedings of the IEEE International Conference
  on Computers, Systems, and Signal Processing}, pp.~175--179, {IEEE} Computer
  Society Press, 1984.

\bibitem{Meyer1999}
D.~A. Meyer, ``Quantum strategies,'' {\em Physical Review Letters}, vol.~82,
  no.~5, p.~1052, 1999.

\bibitem{Eisert1999}
J.~Eisert, M.~Wilkens, and M.~Lewenstein, ``Quantum games and quantum
  strategies,'' {\em Physical Review Letters}, vol.~83, no.~15, p.~3077, 1999.

\bibitem{Andronikos2018}
T.~Andronikos, A.~Sirokofskich, K.~Kastampolidou, M.~Varvouzou, K.~Giannakis,
  and A.~Singh, ``Finite automata capturing winning sequences for all possible
  variants of the {PQ} penny flip game,'' {\em Mathematics}, vol.~6, p.~20, Feb
  2018.

\bibitem{Andronikos2021}
T.~Andronikos and A.~Sirokofskich, ``The connection between the {PQ} penny flip
  game and the dihedral groups,'' {\em Mathematics}, vol.~9, no.~10, p.~1115,
  2021.

\bibitem{Andronikos2022a}
T.~Andronikos, ``Conditions that enable a player to surely win in sequential
  quantum games,'' {\em Quantum Information Processing}, vol.~21, no.~7, 2022.

\bibitem{Giannakis2019}
K.~Giannakis, G.~Theocharopoulou, C.~Papalitsas, S.~Fanarioti, and
  T.~Andronikos, ``Quantum conditional strategies and automata for prisoners'
  dilemmata under the {EWL} scheme,'' {\em Applied Sciences}, vol.~9, p.~2635,
  Jun 2019.

\bibitem{Giannakis2015a}
K.~Giannakis, C.~Papalitsas, K.~Kastampolidou, A.~Singh, and T.~Andronikos,
  ``Dominant strategies of quantum games on quantum periodic automata,'' {\em
  Computation}, vol.~3, pp.~586--599, nov 2015.

\bibitem{Andronikos2022}
T.~Andronikos and M.~Stefanidakis, ``A two-party quantum parliament,'' {\em
  Algorithms}, vol.~15, no.~2, p.~62, 2022.

\bibitem{Theocharopoulou2019}
G.~Theocharopoulou, K.~Giannakis, C.~Papalitsas, S.~Fanarioti, and
  T.~Andronikos, ``Elements of game theory in a bio-inspired model of
  computation,'' in {\em 2019 10th International Conference on Information,
  Intelligence, Systems and Applications ({IISA})}, pp.~1--4, {IEEE}, jul 2019.

\bibitem{Kastampolidou2020a}
K.~Kastampolidou, M.~N. Nikiforos, and T.~Andronikos, ``A brief survey of the
  prisoners' dilemma game and its potential use in biology,'' in {\em Advances
  in Experimental Medicine and Biology}, pp.~315--322, Springer International
  Publishing, 2020.

\bibitem{Kostadimas2021}
D.~Kostadimas, K.~Kastampolidou, and T.~Andronikos, ``Correlation of biological
  and computer viruses through evolutionary game theory,'' in {\em 2021 16th
  International Workshop on Semantic and Social Media Adaptation {\&}
  Personalization ({SMAP})}, {IEEE}, 2021.

\bibitem{Kastampolidou2020}
K.~Kastampolidou and T.~Andronikos, ``A survey of evolutionary games in
  biology,'' in {\em Advances in Experimental Medicine and Biology},
  pp.~253--261, Springer International Publishing, 2020.

\bibitem{Kastampolidou2021}
K.~Kastampolidou and T.~Andronikos, ``Microbes and the games they play,'' in
  {\em {GeNeDis} 2020}, pp.~265--271, Springer International Publishing, 2021.

\bibitem{Kastampolidou2021a}
K.~Kastampolidou and T.~Andronikos, ``Game theory and other unconventional
  approaches to biological systems,'' in {\em Handbook of Computational
  Neurodegeneration}, pp.~1--18, Springer International Publishing, 2021.

\bibitem{Papalitsas2021}
C.~Papalitsas, K.~Kastampolidou, and T.~Andronikos, ``Nature and
  quantum-inspired procedures {\textendash} a short literature review,'' in
  {\em {GeNeDis} 2020}, pp.~129--133, Springer International Publishing, 2021.

\bibitem{Adam2023}
S.~Adam, P.~Karastathis, D.~Kostadimas, K.~Kastampolidou, and T.~Andronikos,
  ``Protein misfolding and neurodegenerative diseases: A game theory
  perspective,'' in {\em Handbook of Computational Neurodegeneration},
  pp.~1--12, Springer International Publishing, 2023.

\bibitem{wootters1982single}
W.~K. Wootters and W.~H. Zurek, ``A single quantum cannot be cloned,'' {\em
  Nature}, vol.~299, no.~5886, pp.~802--803, 1982.

\bibitem{Cruz2019}
D.~Cruz, R.~Fournier, F.~Gremion, A.~Jeannerot, K.~Komagata, T.~Tosic,
  J.~Thiesbrummel, C.~L. Chan, N.~Macris, M.-A. Dupertuis, and
  C.~Javerzac-Galy, ``Efficient quantum algorithms for {GHZ} and w states, and
  implementation on the {IBM} quantum computer,'' {\em Advanced Quantum
  Technologies}, vol.~2, no.~5-6, p.~1900015, 2019.

\bibitem{IBMQuantumComposer2023}
IBM, ``{IBM} {Q}uantum {C}omposer.''
  \url{https://quantum-computing.ibm.com/composer}.
\newblock Accessed: 2022-04-03.

\bibitem{Ampatzis2023}
M.~Ampatzis and T.~Andronikos, ``Quantum secret aggregation utilizing a network
  of agents,'' {\em Cryptography}, vol.~7, no.~1, p.~5, 2023.

\bibitem{Nielsen2010}
M.~A. Nielsen and I.~L. Chuang, {\em Quantum computation and quantum
  information}.
\newblock Cambridge University Press, 2010.

\bibitem{Mermin2007}
N.~Mermin, {\em Quantum Computer Science: An Introduction}.
\newblock Cambridge University Press, 2007.

\bibitem{Qiskit2023}
Qiskit, ``Qiskit open-source quantum development.'' \url{https://qiskit.org}.
\newblock Accessed: 2022-04-03.

\end{thebibliography}

\end{document}